\RequirePackage{fix-cm}
\documentclass[twocolumn,epjc3]{svjour3}  
\smartqed  % flush right qed marks, e.g. at end of proof
\RequirePackage{graphicx}
\usepackage{amssymb}
\usepackage{epstopdf}
\usepackage{amsmath}

\usepackage{subfigure}

\def\s{\sigma}

\def\La{\Lambda}

\def\Ups{\Upsilon}

\def\til{\tilde}

\def\rm{\mathrm}
\def\cal{\mathcal}

\def\be{\begin{equation}}
\def\ee{\end{equation}}
\def\br{\begin{eqnarray}}
\def\er{\end{eqnarray}}
\def\bsub{\begin{subequations}}
\def\esub{\end{subequations}}

\def\til{\tilde}

\def\rm{\mathrm}
\def\cal{\mathcal}

\def\Ups{\Upsilon}

\def\br{\begin{eqnarray}}
\def\er{\end{eqnarray}}
\def\be{\begin{equation}}
\def\ee{\end{equation}}
\def\({\left(}
\def\){\right)}
\def\rlx{\relax\leavevmode}
\def\IR{\rlx\hbox{\rm I\kern-.18em R}}

\def\s{\sigma}

\def\bul{$\bullet$}

\journalname{}
\begin{document}

\title{On the thermodynamics of scale factor dual Universes %\thanksref{t1}
}


\author{G.M. Sotkov\thanksref{e1,addr1}
        \and
        A.L. Alves Lima\thanksref{e2,addr1}
        \and
        U. Camara da Silva\thanksref{e3,addr1}
}

\thankstext{e1}{gsotkov@yahoo.com.br}
\thankstext{e2}{andrealves.fis@gmail.com}
\thankstext{e3}{ulyssescamara@gmail.com}

\institute{Departamento de F\'\i sica - CCE\\
Universidade Federal de Espirito Santo\\
29075-900, Vitoria - ES, Brazil \label{addr1}
         }

\date{%Received: date / Accepted: date
}

\maketitle

\begin{abstract}
The thermodynamical aspects of the conformal time scale factor duality (SFD)  of cosmological models  within  Einstein Gravity 
are investigated. We  derive the  SFD  transformations  of the  thermodynamical quantities describing the thermal evolution of the matter fluid 
and of the apparent horizon.
The thermodynamical properties of the self-dual cosmological models with a modified Chaplygin gas  
are studied in detail. We deduce the restrictions on the equation of state parameters that allow to extend  scale factor duality as a UV/IR symmetry of the cosmological models consistent with their thermodynamical behavior.  
\end{abstract}

\section{Introduction}
\label{intro}
Although the effective field theories  used in the description of the universe evolution contain a few distinct energy (or length) scales and other dimensionful parameters,  it is expected  that local conformal  transformations take place  as an \emph{asymptotic} symmetry \cite{motolla,antoniadis2012conformal,khoury1,kaloper,thooft,bars}. There are a variety of options for dynamical and spontaneous breaking of the conformal (Weyl) symmetry which lead to rather realistic cosmological models, consistent with recent astrophysical observations  \cite{planck}.

The \emph{scale factor duality} (SFD) invariant cosmologies \cite{venez,VezianoPrebigbangincosmo,dabrow,chim-wz,Chimento:2008qq,camara}, for Friedmann-Roberton-Walker (FRW) universes\footnote{We denote $\kappa^2 \equiv 16\pi G$, $f' \equiv \frac{df}{d\eta}$ a derivative with respect to conformal time $\eta$, and $k=0,\pm 1$ for spatially flat, open or closed universes.}
\begin{eqnarray}
ds^2\!=\!a^2(\eta)\!\left( -d\eta^2 \!+\! \frac{dr^2}{1-k r^2} \!+\!  r^2 d\theta^2 \!+\! r^2 \sin^2\theta \, d\varphi^2 \right), \label{frwmetric}
 \end{eqnarray}
 provide   examples  of models  with a manifest  residual $Z_2$ symmetry, where  the Weyl transformations are partially broken while preserving the invariance under the subgroup of scale factor inversions, $a \mapsto 1/a$.
They represent a symmetry of the  space of solutions of  the  Friedmann  equations, 
\begin{eqnarray} 
 &&\rho'+3\frac{a'}{a}(\rho+p)=0, \ \frac{\kappa^2}{6} \,\rho = \left(\frac{a'}{a^2}\right)^2+ \frac{k}{a^2},\nonumber\\
 &&\frac{\kappa^2}{2} \, p = \left(\frac{a'}{a^2}\right)^2 - \frac{2a''}{a^3} - \frac{k}{a^2}, \label{frw}
\end{eqnarray} 
interchanging  small and large physical scales. 
 When combined with time reflections, they give rise to scale factor duality transformations  known to be an important tool in the construction of pre-big-bang cosmological models \cite{venez,VezianoPrebigbangincosmo}.  In the case when the SFD requirement is implemented in \emph{conformal time}, combined with the specific transformation of the matter fluid with equation of state (EoS)  $p / \rho = \omega(\rho)$ \cite{camara,ds2016scale},
\br
&&\tilde{a}(\tilde{\eta})= c_0^2 / a(\eta), \  \til{\eta}=\pm \eta + {\text{const.}},\nonumber\\
&&\til {a}^2 \til{\rho}(\til a)= a^2 \rho(a),\nonumber\\
&&\til {a}^2 [3\til {p}(\til a)+ \til{\rho}(\til a)] \!=\! -a^2 [3p(a)+\rho(a)],\nonumber\\
&&\omega (\rho)+\til{\omega}(\til{\rho})\!=\! -\frac{2}{3}, \label{sfda}
\er
one observes that, apart from  keeping  invariant the form of the Friedmann equations (\ref{frw}), they  also manifest  an interesting \emph{UV/IR}  feature. Namely, for  fluids satisfying the null energy condition (NEC), i.e. when $-1\leq \omega \leq\frac{1}{3}$ (and the same for its dual $\til{\omega}$), \emph{high energy} densities $\rho$ and large values of the scalar Ricci curvature are always mapped into \emph{small} corresponding values in the dual models. 
%As we have recently demonstrated \cite{sfd-ccc}, there are a few distinct realizations%
%
%\footnote{When we consider cosmological models with a finite ``conformal lifetime'' (i.e. $0 \leq \eta \leq \eta_f$) undergoing one period of decelerated and one of accelerated  expansion (or contraction).}  
%
%of the conformal time SFD (\ref{sfda}), depending on the particular choice  of the conformal time transformations.
We shall consider universes with one period of decelerated and one period of accelerated expansion, such that the ``conformal lifetime'' is finite; say $0 \leq |\eta| \leq \eta_f$ as $0 \leq a(\eta) \leq \infty$. Then, as we have recently demonstrated \cite{ds2016scale}, there are a few distinct realizations of the conformal time SFD (\ref{sfda}), depending on the particular choice  of the conformal time transformations:
 (i)  $\til \eta= - \eta$  gives rise to an expansion/expansion type of pre-big-bang and conformal cyclic cosmologies; (ii) $\til \eta =\eta+\eta_f$  leads to contraction/expansion pre-big-bang models and (iii) $\til \eta=2\eta_c-\eta$  defines a family of SFD self-dual post-big-bang cosmologies \cite{camara}, with  $\eta_c=\eta_f /2$ being the instant of deceleration/acceleration transition, i.e. $\ddot a(\eta_c)=0$. 
The  investigation of the consequences of the \emph{self-duality} requirement \cite{camara} has revealed a new application of  SFD out of the scope of the original pre-big-bang scenario \cite{venez,VezianoPrebigbangincosmo}---it allows one to describe the short distance and early time (UV) behavior of the universe evolution in terms of its large distance and late times (IR) behavior.

The present paper is devoted to the  investigation of the thermodynamical (TD) properties of the self-dual cosmological models  introduced in ref.\cite{camara}. 
 The problem we address here concerns the conditions  that permit  one to extend the UV/IR symmetry of the curvature and energy density to the thermodynamical characteristics of the \emph{SFD symmetric universes} as well. More precisely, our aim is to derive the SFD transformations of the temperature, entropy, internal energy, etc., and to further  select the values of the equation of state  parameters  of these models such that  high temperatures are transformed into low temperatures  and vice-versa. 
As expected, the self-duality of the FRW equations  also  allows us to demonstrate the self-duality of the thermodynamical description of the corresponding pairs of \emph{dual apparent horizons} within the framework of the Cai-Kim prescription for their temperature  and entropy \cite{caikim}.   It  deserves to highlight  one interesting result of our study of  the thermodynamical  features of the simplest \emph{self-dual fluids}, based on the radiation-like Chaplygin gas with EoS 
$$p=\frac{1}{3} \rho-\frac{4}{3} \rho^{\delta}_{\Lambda}\rho^{1-\delta} $$
namely that 
 within the interval $\delta \in[\frac{3}{4},1]$  not only the curvature and matter density,  but also the temperature and the pressure turn out to be monotonically decreasing with the increasing of the scale factor. As a consequence the  \emph{UV/IR symmetric}  ``thermal history''  of these self-dual cosmological models appears to be  quite similar to the standard $\Lambda$CDM model. 

\section{Thermodynamical aspects of scale factor duality}

Given the conformal time SFD transformation laws (\ref{sfda}),
%with $\til \eta =2\eta_c-\eta$
 we are interested in deriving the  transformations  of all the remaining thermodynamical  characteristics  of  a barotropic  fluid  within  a constant comoving volume $V_0$. 
The  thermodynamical behavior of such fluids is described by an adiabatic  process in thermal equilibrium (see, e.g. \cite{weinberg}),
%with constant comoving entropy $\cal S \equiv s \, a^3$ and
with the entropy, energy and pressure densities $s$, $\rho$ and $p$, respectively, all functions of temperature $T$ only.
Then  as a consequence of the first TD law, $dE=TdS - pdV$, one gets the well known expressions
%for total internal energy $E \equiv \rho \, V$,  entropy $S \equiv s \, V$ and (physical) volume $V  = V_0 \, a^3$,   
\begin{eqnarray}
s=\frac{p+\rho}{T} ,\ \ \frac{d s}{d T} =\frac{1}{T} \;  \frac{d \rho }{d T}, \ \
 \frac{d\rho}{dT}=\frac{1}{T}\left(p+\rho\right)\frac{d\rho}{dp}. \label{TDcons}
\end{eqnarray}
for the total internal energy $E \equiv \rho \, V$,  the entropy $S \equiv s \, V$ and the physical volume $V  = V_0 \, a^3$.
%of the considered fluids. 
Since the entropy is time independent  $d S / d t = 0$,  the entropy density scales as $s = \cal S / a^3$ for a constant $\cal S$.
From the simple form of the above identities, one may easily derive  the  SFD transformations  of  temperature, entropy  and  internal energy:
\br
\frac{\til E}{\til a}\!=\!\frac{E}{a},\ \til S =S, \ \frac{1}{\til a}\left(\til S\til T \!-\! \frac{2}{3} \til E\right)=\! - \frac{1}{a}\left(ST \!-\! \frac{2}{3} E \right), \label{tdsfd}
\er 
by taking into account  Eq.(\ref{sfda}). This establishes the desired SFD relations between the thermodynamics of a pair of \emph{dual} fluids.

\subsection{Self-Dual Fluids Thermodynamics}	\label{SectSFDfluidThermo}

According to the results of ref.\cite{camara}, by taking a pair of scale factor dual FRW solutions  $( a ,\rho,  p)$ and  $(\til a,\til \rho, \til p )$ and requiring their  \emph{self-duality}:
%one can   construct a special class of  \emph{ post-big-bang} cosmologies, representing  a \emph{single SFD symmetric} two periods (decelerated and accelerated)   expanding Universe  that satisfies the  following \emph{self-duality} condition:
\be 
%\tilde{a}(\eta)=a(\tilde{\eta}),\quad\quad 
\til{\rho}(\til a)= \rho(\til a),\ \til{p}(\til a)= p(\til a), \ \til{\omega}(\til {\rho})=\omega(\til{\rho}),\ \tilde \eta = 2 \eta_c - \eta,
\label{sfsd1}
\ee
one may obtain 
a special class of  (post-big-bang) cosmologies in which SFD is a symmetry of one \emph{single} FRW universe with two periods of acceleration. That is, SFD maps  accelerated into decelerated phase---thus the early universe into the late universe---, through a reflection about the SFD-invariant  instant $\eta_c$.
% when acceleration vanishes.

The symmetry requirements (\ref{sfsd1}) impose  quite strong restrictions on the density of a self-dual fluid. Combining (\ref{sfsd1}) and (\ref{sfda}) we find that its EoS must obey 
%be such that 
\br
&&\frac{c_0^2}{a(\eta)}=a(2\eta_c - \eta)  , \quad \rho( {\Omega} a)= {\Omega}^{-2}\rho(a),\nonumber\\
&&\omega(\rho)+ \omega({\Omega}^{-2}\rho)= -\frac{2}{3}, \ \Omega\equiv\frac{c_0^2}{a^2} .
\label{homog} 
\er
Within a vast family of two-component interacting  fluids, the ``radiation-like'' modified Chaplygin gas models \cite{chap-modif1,Benaoum:hh,chap-mod2}, 
\br
\rho=\Big( \rho_\La^\delta +  \rho_r^\delta \; a^{-4\delta}\Big)^{\frac{1}{\delta}},\ \ p=\tfrac{1}{3} \rho -\tfrac{4\rho_\La^\delta}{3}  \;  \rho^{1-\delta},
\label{chaply}
\er
are the only ones to satisfy these constraints, as long as  $\til \rho_r=\rho_r=c^4_0 \rho_{\Lambda}$ and $ \til \rho{_\Lambda}=\rho_{\Lambda}$. In the case of $\delta>0$, with $\rho_{\Lambda}\leq\rho<\infty$, they provide a class of self-dual, $\Lambda$CDM-like big-bang cosmologies \cite{camara}. According to Eq.(\ref{chaply}), the self-dual fluid  behaves as radiation when $a\rightarrow 0$, and  as a cosmological constant $\rho_{\Lambda}$ when $a \rightarrow\infty$. Thus the big-bang singularity is followed by a decelerated  and then an accelerated  period of expansion within a finite  conformal lifetime whose duration is given by (see ref.\cite{camara})
\br
 \eta_f=2\eta_c=\frac{\sqrt{3}}{4\delta \, (\rho_r \rho_\La)^{\frac{1}{4}}} \frac{\left[\Gamma\left(\frac{1}{4\delta}\right)\right]^2}{\Gamma\left(\frac{1}{2\delta}\right)} . %\quad\quad  \rho_{\Lambda}\leq\rho<\infty 
\nonumber %\label{aetafal}
\er
The transition between the deceleration and acceleration epochs occurs at the instant $\eta_c$,  corresponding to  the SFD invariant value of the scale factor $a(\eta_c) =c_0= \left( \rho_r /\rho_\La\right)^{\frac{1}{4}}$, when $ \rho_0 =\rho(\eta_c) = 2  \rho_\La $ and the deceleration parameter $q(a)\equiv\frac{1}{2}\left(1+3\frac{p}{\rho}\right)$ vanishes. 
% $q(a_{cr})=0$.
We should mention the rather different nature of  the solutions with $\delta<0$  --- their curvature is monotonically decreasing and bounded,  $R_{dS}\geq R \geq 0$.  They describe  \emph{eternally} expanding \emph{non-singular} universes, which  are asymptotically de Sitter at the remote past $\eta\rightarrow -\infty$  and  accelerated up to the moment $\eta_c$. Their  late time decelerated phase is dominated by radiation in the far future, $\eta\rightarrow \infty$. 
%Their curvature is monotonically decreasing and bounded  $R_{dS}\geq R \geq 0$. 
%and in the far future $\eta\rightarrow \infty$ their behavior  is dominated by radiation.
%, i.e. $a(\eta)^2\sim\eta^2$. 

We next consider the thermodynamics of the self-dual fluids (\ref{chaply}). The  evolution of their temperature $T(a)$ as a function of the scale factor or, equivalently, its dependence  on the physical volume  $V=V_0 a^3$, can be  easily derived  from the last of the Eqs.(\ref{TDcons}), once the EoS of the fluid is given: 
\begin{eqnarray}
&&4\,\sigma_{sb}  T^4=\rho\left(1-\frac{\rho^{\delta}_{\Lambda}}{\rho^{\delta}}\right)^{4-\frac{3}{\delta}}= \frac{\rho_r}{a^4} \left(1+\frac{\rho^{\delta}_{\Lambda}}{\rho^{\delta}_r} a^{4\delta}\right)^{\frac{4}{\delta}-4},\  \nonumber\\
&&\s_{sb}=\frac{\pi^2 K^4_B}{60\hbar^3}, 
%\quad T(a)= \frac{3}{4} \frac{\rho^{\delta}_r}{\mathcal{S}} a^{3-4\delta} \rho^{1-\delta},
\label{temp}
\end{eqnarray}
%obtained from the last of the Eqs.(\ref{TDcons}). 
where we have fixed the constant of integration to be equal to the Stephan-Boltzmann constant  $\s_{sb}$.
%=\tfrac{\pi^2 K^4_B}{60\hbar^3}$.
Such a choice reproduces, in the limit $\rho_{\Lambda}\rightarrow0$,%
\footnote{\label{footlimiti}I.e. when (\ref{chaply}) reduces to pure radiation, with $p = \tfrac{1}{3} \rho$ and $\rho = \rho_r / a^4$.} 
the Stephan-Boltzmann law $\rho=4\, \s_{sb} \,T^4$  for an ultrarelativistic gas. 
Notice, however, that  the thermal evolution of the scale factor $a(T)$ and of the fluid density  $\rho(T)$ can be explicitly found  only in a few cases (say for  $\delta=1$ and $\delta=3/4$),  when the inversion of Eqs.(\ref{temp}) is available. Nevertheless, the above equations will  allow us to  deduce  their TD  properties  in general,  and also to establish the restrictions on the values of  EoS  parameter $\delta$ that ensure their physical consistency. 

Let us remind one important feature of the equilibrium  processes (in homogeneous systems), namely that  all the thermodynamical potentials---the internal energy $E(S,V)$, the free energy $F(T,V)=E-TS$, the Gibbs potential $\Phi(T,P)=E+pV-ST$, etc.---have to remain at their minimal values. The  conditions to guarantee the stability of these minima, say, against thermal or pressure  fluctuations,  are given by the so called \emph{thermodynamic inequalities} (see for example \cite{landau-v5}).  In the case of the adiabatic  equilibrium processes occurring  in the  self-dual fluids (\ref{chaply}) it is sufficient to impose that its adiabatic compressibility is always positive, i.e.
$$K_S= - \frac{1}{V}\left(\frac{\partial V}{\partial p}\right)_S > 0,\ {\text{or equivalently}} \ \frac{dp}{d\rho}> 0.  $$  
In order to establish the  consequences of such  a ``TD stability restriction''  we have to  examine  the behavior of the temperature $T(\rho)$  and the pressure $p(\rho)$ as functions of  the fluid density. Observe that  \emph{they are not  monotonic functions} for all the values of $\delta$,%
\footnote{Let us remind, however, that $\rho(a)$  \emph{is}  always monotonically decreasing  for all $a\in(0,\infty)$.}
due to the  existence of real zeros $\rho_{cr}(\delta)= \left(4-4\delta\right)^{\tfrac{1}{\delta}} \rho_{\Lambda}$ of the derivatives $\frac{dp}{d\rho}$ and $\frac{dT}{d\rho}$, placed within the intervals  $\rho_{cr}\in(\infty,\rho_{\Lambda}]$ when  $0<\delta\leq3/4$, or  $\rho_{cr}\in[\rho_{\Lambda}, 0]$ for  $\delta<0$.
The corresponding ``critical values"
\begin{eqnarray}
&&p_{cr}(\delta)=-\frac{4}{3}\delta \rho_{\Lambda} \left(4-4\delta\right)^{\tfrac{1-\delta}{\delta}},\nonumber\\
&&T_{cr}=T_* \left(4-4\delta\right)^{\tfrac{1-\delta}{\delta}} \left(3-4\delta\right)^{\tfrac{4\delta-3}{4\delta}},\nonumber\\
&&T_* \equiv\left(\frac{\rho_{\Lambda}}{4\s_{sb}}\right)^{\tfrac{1}{4}} \label{critic}
\end{eqnarray}
represent the maxima of $p(\rho)$ and $T(\rho)$ for $\delta<0$ and the minima for $0<\delta<3/4$. 
Notice that the $\delta = 3/4$ model admits a minimum  for the pressure only. One can also easily derive  the  specific \emph{scaling} behavior of all the TD quantities around these critical points (for $\delta\leq 3/4$), viz.
\begin{eqnarray}
&&K_S \sim|T-T_{cr}|^{-\tfrac{1}{2}},\nonumber\\ 
&&\rho-\rho_{cr}\sim |T-T_{cr}|^{\tfrac{1}{2}},\nonumber\\
&&p-p_{cr}\sim(\rho-\rho_c)^2,\label{critexp}
\end{eqnarray}
etc., with universal critical exponents independent of the values of $\delta$.  Although the second derivatives of the TD potentials manifest power-like singularities, typical for  second order phase transitions,  we should emphasize however that the above described critical behavior \emph{does not satisfy} all the requirements \cite{landau-v5} needed for the realization of such phase transitions.
%, i.e. $\rho_{cr}$ \emph{is not an extremum}  of the temperature. 

The above discussion makes it clear  that  the fluids with
 \br
  &&(i) \ \delta<0,\ \ (ii)\ 0<\delta <3/4,\nonumber\\
  &&(iii)\ \delta=3/4,\ (iv)\quad  \delta>3/4 	\label{cases}
 \er
 should represent different thermodynamical properties.
  % Another consequence of the  specific form (\ref{chaply}) of the considered self-dual EoS is related to the  conditions imposed on $\delta$  by the requirement on the values of the sound velocity $0\leq v^2_s\leq 1$ and related TD instabilities.
In case \textit{(ii)} we realize that $K_S$ changes its sign at $\rho_{cr}$ and  the Chaplygin-like fluids, filling the expanding asymptotically de Sitter universe, manifest a rather \emph{unphysical  TD behavior}---after reaching its minimal value $T_{cr}$, the temperature of the expanding fluid 
starts \emph{increasing} towards infinity, so the fluid becomes ``extremely rarefied and infinitely hot''. 
 
The  thermal history of the eternal universe in case \emph{(i)}, when $\delta<0$, 
has a cooler,  much more reasonable,  but  still  \emph{ unstable}  behavior,  since again the TD inequality $K_S>0$ is not respected  during all of its evolution.  Both the initial de Sitter state, and the final radiation stage have a vanishing temperature. The transition between these two cold phases represents a ``heating-to-cooling'' process, with the temperature increasing from zero up to $T_{cr}(\delta<0) = T_{max}$ before decreasing to zero again. It is worth reminding that for small enough values of $|\delta|$ this models  describe a  hilltop inflationary period (in  the neighborhood of initial de Sitter state), that turns out to fit quite well with the Planck-2015 data \cite{planck}, cf.  ref.\cite{camara}.  

The models with $\delta >\frac{3}{4}$  are examples of stable TD behavior, since $K_S$ remains always positive,  and   their TD evolution  is described by monotonically decreasing temperature $T(\rho)\in(\infty, 0)$ and pressure $p\in(\infty, -\rho_{\Lambda})$, as in most  of the $\Lambda$CDM-like models.  We shall further impose a natural upper bound $\delta\leq1$ that excludes the models whose (Ricci) curvature $ R=2 \rho^{\delta}_{\Lambda} \left( \rho \right)^{1-\delta}$  increases with the decreasing of the fluid density.

%----------------- BEGIN FIGURE 
\begin{figure}[htbp]
\begin{center}
\includegraphics[scale=0.9]{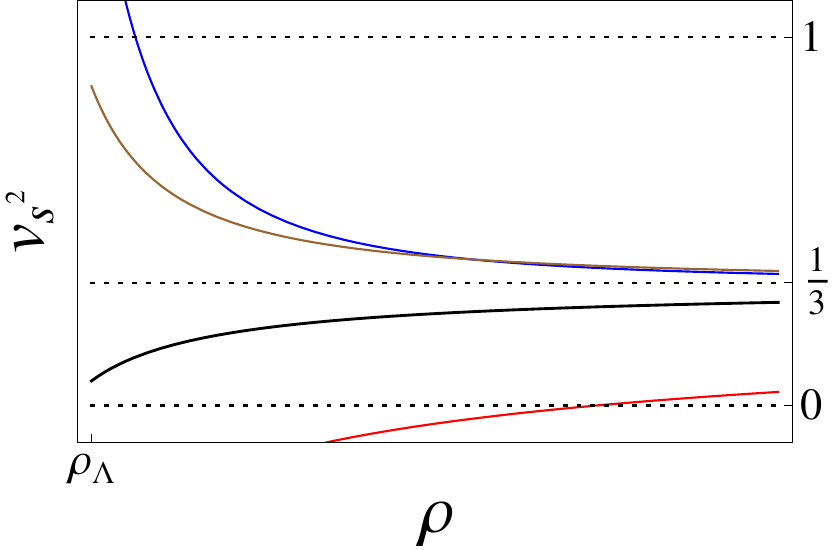}  
\caption{
Behavior of the speed of sound for different values of $\delta$. Blue: $\delta > \frac{3}{2}$; Brown: $1 < \delta < \frac{3}{2}$; Black: $\frac{3}{4} < \delta < 1$; Red: $0 < \delta < \frac{3}{4}$.
}
\label{SoundSpeed}
\end{center}
\end{figure}

%---------------- END FIGURE

Finally, the  thermodynamics of the $\delta=3/4$ model \emph{(iii)} is quite different from  all  the other cases we have described above. Here,  $\rho_{cr} = \rho_\Lambda$ is \emph{not} an extremum  of the temperature, which is a monotonic function. So this is the only asymptotically  de Sitter self-dual model  with a \emph{non-vanishing} final temperature $T_*$, which nevertheless shares (together with the simplest $\delta=1$ model) the standard radiation-like Stephan-Boltzmann law $\rho=4\,\sigma_{sb}  T^4$.
The pressure reaches its minimum $p_{cr} = -\rho_{\Lambda}$ only at the final de Sitter boundary, when $a \to \infty$ and $\rho \to \rho_{cr} = \rho_\La$. 
%In this limit, the adiabatic  compressibility $K_S=\frac{1}{\rho}\left(\frac{\partial \rho}{\partial p}\right)_S \approx \frac{4}{\rho-\rho_{cr}} + {\rm{const.}}$ has a  power-law singularity.  This may be interpreted as a second order phase transition at the end of the universe evolution. One can further  suggests that the final de Sitter boundary of the  $\delta=3/4$ self-dual cosmological model  may be described as an  euclidean three dimensional CFT related to the critical behavior of the fluid. 
 It is worthwhile to mention that such $\delta = \frac{3}{4}$ self-dual universe exhibits many of the features of the Chaplygin-like \emph{quintessence} models \cite{Kamenshchik:2001ec,chap-modif1,Benaoum:hh}. Certain similarities with the standard $\Lambda$CDM cosmology should be highlighted  as well \cite{camara}---its energy density  at early times is approximated by radiation, $\rho \approx \rho_{r}/a^4$, while at relatively late times it behaves as a cosmological constant with cold dark matter, i.e. $\rho \approx \rho_{\Lambda} +\rho_{dm}/a^3$.

\vspace{3mm}

Another set of  restrictions on the values of  the parameter $\delta$ arises from the requirement that  the speed of sound $v^2_s=\tfrac{dp}{d\rho}$ should be smaller than $c^2 = 1$, i.e.
$0\leq v^2_s\leq 1$, which imposes 
\br
0\leq v_s^2 = \frac{1}{3} - \frac{4(1-\delta)}{3} \frac{\rho^{\delta}_{\Lambda}}{\rho^{\delta}}\leq 1 \label{soundv}
\er
 for the modified Chaplygin gas models. 
 %The behavior of $v_s^2(\rho)$ is depicted in Fig.\ref{SoundSpeed} for different values of $\delta$. 
 Since $v^2_s(\rho)$ turns out to be a monotonic function, it is enough to consider the consequences of Eq.(\ref{soundv}) for its initial and final values $v^2_{s(in)} \equiv v_s^2(a = 0)$ and $v^2_{s(f)} \equiv v_s^2(a \to \infty)$. They are
\begin{eqnarray}
v^2_{s(in)}&=&-1-\frac{4}{3}|\delta| <0, \ v^2_{s(f)} =\frac{1}{3}, \ {\text{for}} \ \delta<0,\nonumber\\
v^2_{s(in)}&=&\frac{1}{3}, \  v^2_{s(f)}= -1 +\frac{4}{3}\delta, \ {\text{for}} \ \delta>0.
\label{pmsoundv}
\end{eqnarray}
This makes it evident which are the selected ranges of values of $\delta$ that provide \emph{hydrodynamically consistent} equations of state:

\bul \quad Models with $\delta>\tfrac{3}{2}$ admit sound waves with superluminal velocity $v^2_s>1$ (blue line in Fig.\ref{SoundSpeed}), and hence  must be discarded.

\bul \quad The only self-dual fluids that  satisfy  
% the sound velocity conditions 
(\ref{soundv}) are those with $\delta\in[\tfrac{3}{4}, \tfrac{ 3}{2}]$. 
Notice, however, that at $\delta=1$ there occurs an important change in the  behavior of $v_s(\rho)$. We have
\br
\tfrac{1}{3} \ge v_s^2(\rho)\ge v^2_s(\rho_{\Lambda})\ge 0 &\quad {\text{for}} \quad  \tfrac{3}{4}\le\delta\le1 ; \nonumber \\
\tfrac{1}{3} < v_s^2(\rho)\le v^2_s(\rho_{\Lambda})\le 1 &\quad  {\text{for}} \quad 1< \delta \le \tfrac{3}{2}. \nonumber
\er
(Black and brown lines in Fig.\ref{SoundSpeed}, respectively.)
While, for $\delta < 1$, $v_s^2$ is a monotonic function decreasing with the density, for  $\delta>1$ it becomes a function which \emph{increases} as the density decreases. Note that the latter is a rather \emph{unphysical} behavior.

\bul\quad  For all the cosmological models with $\delta<\tfrac{3}{4}$, the lower bound is \emph{not respected}, i.e. we have  $v^2_s<0$ for a certain range of values of the fluid density (red line in Fig.\ref{SoundSpeed}), which causes  thermo- and hydrodynamical  instabilities when the zero sound velocity limit is violated.

\vspace{3mm}

As we have already mentioned,  the family of  models with $\delta <\tfrac{3}{4}$ is in fact separated in two distinct classes  --- one for $\delta<0$ and the other for $\delta\in(0,\tfrac{3}{4})$ --- with quite different geometrical and thermodynamical properties.  These differences are highlighted in the description of the fluid as a self-interacting scalar field. The equation of state (\ref{chaply}) is equivalent to the following  scalar field potential \cite{camara} 
%In order to highlight the nature of these differences we find it worthwhile  to remind the equivalent self-interacting scalar field description of the considered self-dual fluids \cite{camara} 
\begin{eqnarray} 
V(\sigma)=&&\frac{2}{\kappa^2 L^2}\Bigg\{\left[\cosh^2\left(\frac{\delta}{\sqrt{2}}\kappa\sigma\right)\right]^{\frac{1}{\delta}}+\nonumber\\
&&2 \left[\cosh^2\left(\frac{\delta}{\sqrt{2}}\kappa\sigma\right)\right]^{\frac{1-\delta}{\delta}}\Bigg\},\quad L^2 \equiv\frac{2}{\rho_{\Lambda}} .  \label{poten}
\end{eqnarray}
As one can easily verify, the relation between the EoS parameter   $\delta$ and the scalar field mass $m^2_{\s}=V''(0)$ at  the ``de Sitter extremum" $\s=0$ of the potential $V(\s)$, 
\br
m^2_{\s}L^2=2 \delta \left(3 - 2\delta \right),\label{dimension}
\er
provides a (partial) explanation  of the qualitatively different behavior of the models corresponding to the discussed  different ranges of values for $\delta$. For example, in the  cases of $\delta<0$ and $\delta>\tfrac{3}{2}$  the corresponding potentials both have  maxima at $\s=0$, due to a negative mass  squared  $m^2_{\s}<0$, which  is in the origin of their  %nearly de Sitter 
instabilities.
 On the other hand, the cosmological models with $0< \delta \leq \frac{3}{2}$  have $m^2_{\s}\in[0,\tfrac{9}{4}]$, and the scalar field potential (\ref{poten}) has  a parabolic-like shape with a global minimum at $\s=0$. 
 Still, the models with $\delta\in(0,\tfrac{3}{4})$ and $\delta\in[\tfrac{3}{4},\tfrac{3}{2}]$ represent two qualitatively different thermal histories of the universe evolution, reflecting  the change in the TD properties of the fluid that occurs at $\delta=\tfrac{3}{4}$. Let us  also remind that  the restriction $\delta\leq1$ (as we have demonstrated above) is of a purely gravitational and hydrodynamical nature.

\vspace{3 mm}

Our brief discussion of the thermodynamical and hydrodynamical properties  of the self-dual cosmological models  (\ref{chaply}) has pointed out a set of  arguments in favor of the  physical consistency of the  models with  $\tfrac{3}{4}\leq\delta\leq1$.  It remains, however, to verify  whether the conformal time SFD does act as thermodynamical UV/IR symmetry for these selected models.

\subsection{SFD as  UV/IR symmetry}

The SFD transformations for the thermodynamical characteristics of  modified Chaplygin gas follow the general equations (\ref{tdsfd}). The constant entropy can be fixed by taking the radiation limit,%
\footnote{I.e. when $a^4 \ll \rho_r / \rho_\La$; see footnote \ref{footlimiti} above.}
as
$$\til S = S = \frac{4}{3} V_0\left(4\sigma_{sb} \rho_r^3\right)^{1/4},$$ and the ``fixed point'' of the physical volume transformation $\til V = v_0^2 / V$  is found to be $v_0 \equiv V_0 \left( \rho_r /\rho_\La\right)^{\frac{3}{4}}$.
%, such that $\til V = v_0^2 / V$.
Now the transformations of the temperature and of the entropy density $s(a)$, with the aid of Eqs.(\ref{temp}) and (\ref{sfda}), can be written rather simply as
\br
\til {T}(\til {a})\!=\!\Omega^{1-2\delta}T(a), \ \tilde s (\tilde a)\!=\! \frac{S^2}{v^2_0} \frac{1} {s (a)},\ 
{\text{where}} \ \Omega \!\equiv\! \frac{c_0^2}{a^2} .		\label{temptrans}
\er
Immediately we see that the fluid with $\delta=1/2$ has the interesting property of its temperature being invariant under the scale factor duality, $\til {T}(\til {a})=T(a)$, and thus $T_{in}=\infty=T_{f}$. For all  other values of $\delta$, the 
temperature transformation  $\til{T}(T)$  is given by 
\br
\left(\til T T\right)^{\tfrac{1}{4\delta-2}}=T_* \left( T^{\tfrac{2\delta}{2\delta-1}} + \til T^{\tfrac{2\delta}{2\delta-1}} \right)^{\tfrac{1-\delta}{\delta}},		 \label{TT}
\er
obtained by substituting the scale factor $(a/c_0 )^{2-4\delta}=\til T/T$  from Eqs.(\ref{temptrans})  into Eqs.(\ref{temp}). The \emph{fixed point} of this transformation is given by the SFD invariant temperature 
\br
T_0=T(\eta_c)=\tilde T(\eta_c)=2^{\tfrac{1-\delta}{\delta}} T_*	\label{invT0}
\er
of the fluid at the moment $\eta_c$ of the transition from decelerated to accelerated expansion, i.e. $T_0$ corresponds to $q(\eta_c)=0$. 
At this same instant, the physical volume of the fluid reaches the invariant value $v_0$, and the EoS parameter takes the particular value $\omega(\eta_c) = -\tfrac{1}{3}$, corresponding to a cosmic string gas --- this is the only value of $\omega$ invariant under the last of Eqs.(\ref{sfda}).

%----------------- BEGIN FIGURE 
\begin{figure}[htbp]
\begin{center}
\includegraphics[scale=0.7]{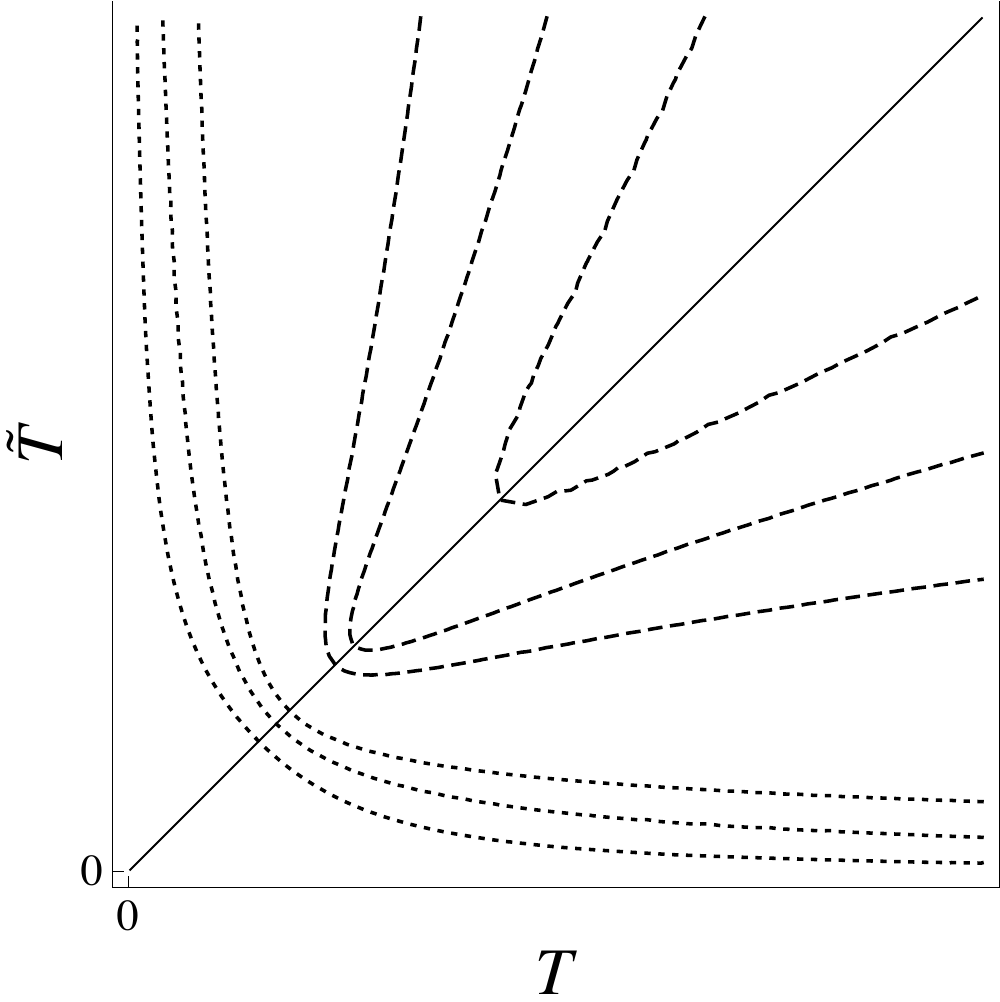}  
\caption{
Dual temperatures related by Eq.(\ref{TT}), for $\delta > 0$. Dashed lines correspond to $0 < \delta < \frac{3}{4}$ and dotted lines to $\delta > \frac{3}{4}$. The diagonal line corresponds to the limiting case $\delta = \frac{1}{2}$.
}
\label{TtildeTfig}
\end{center}
\end{figure}

%---------------- END FIGURE

We next consider the conditions under which the above transformation  $\til T(T)$ acts as a  high-to-low temperature UV/IR symmetry.   
Since (\ref{TT}) gives the explicit action of SFD on the temperature of the self-dual fluid, it maps `initial' temperatures (at small values of $a(\eta)$) to `final' temperatures (at large values of $a(\eta)$). 
%It is not, however, equivalent to say that it maps high temperatures into small ones --- i.e. that it gives is a \emph{`thermodynamic UV/IR symmetry'}.
But this evolution, as described in Sect.\ref{SectSFDfluidThermo},  depends qualitatively on the values of $\delta$ listed in cases (\ref{cases}), so the different behaviors must be described by (\ref{TT}). 
%The UV/IR symmetry will only be present when $T$ ranges from small to large values along the evolution of the universe and this was seen to only happen in case \emph{(iv)}, i.e. for $\delta > 3/4$. 
Although the latter is a complicated formula,  the limit $T\rightarrow \infty$ gives the most relevant information about the nature of the symmetry. For $\delta > 1/2$ we find a simple asymptotic relation between the pair of dual temperatures,
\be
\til T\approx T_*\left(\frac{T_*}{T}\right)^{\frac{1}{4\delta-3}}  ,	\label{TTasymp}
\ee
which confirms for $\delta>\tfrac{3}{4}$ the expected mapping between $T\approx\infty$ and $\til T\approx0$ and vice-versa. 
 The corresponding monotonic curves of $\tilde T(T)$ are depicted in Fig.\ref{TtildeTfig} as dotted lines.
Note that  in the particular case of $\delta=1$ both the asymptotic formula (\ref{TTasymp})  and the exact one, Eq.(\ref{TT}), coincide: $\til T=\tfrac{T^2_*}{T}$. As anticipated in  Sect.\ref{SectSFDfluidThermo}, the  $\delta=\tfrac{3}{4}$  model represents a notable exception  ---  the initial temperature $T\approx\infty$ is now  transformed  into a \emph{finite}  final value  $\til T\approx T_*$.
%, as expected from the results presented in  Sect.\ref{SectSFDfluidThermo}.
 %The monotonic curves of $\tilde T(T)$ are depicted in Fig.\ref{TtildeTfig} as dotted lines.
The  evolution  of the temperature $T(\rho)$  in the models with $0 < \delta < \tfrac{3}{4}$
is  \emph{non-monotonic}, with a minimum value $T_{cr}$.   Consequently, $\tilde T(T)$ is not a single valued function, which gives the dashed lines in Fig.\ref{TtildeTfig}. The ``tip'' of these curves (where they intersect the diagonal line) correspond to the invariant temperature (\ref{invT0}), 
$$T_0 = (2 - 2\delta)^{\frac{\delta -1}{\delta}} \left(3-4\delta\right)^{\tfrac{3-4\delta}{4\delta}} T_{cr},$$ 
cf. Eq.(\ref{critic}).
As expected, since both initial and final temperatures diverge in this case, 
Eq.(\ref{TT}) indeed maps $T\approx\infty$ to $\til T\approx\infty$. 
Finally, for $\delta < 0$ the temperature is bounded, $0\leq{T}\leq{T_{cr}}\equiv{T_{max}}$, and Eq.(\ref{TTasymp}) is not valid. In this case we should consider instead the limit $T \to 0$ of Eq.(\ref{TT}), and a similar  asymptotic expansion of $\til T(T)$ confirms that the SFD dual of $T\approx0$ is indeed $\til T\approx0$, corresponding to the initial and final states being equally cold.

The results of our investigation of the SFD temperature $\til{T}(T)$ transformations (\ref{TT}) may be summarized in the following \emph{statement}:  The relevant physical quantities, $\rho(a)$, $R(a)$, $p(a)$ and $T(a)$, describing  the evolution of the considered  self-dual cosmological models (\ref{chaply})  \emph{are, all, monotonic functions}  only for $\delta\geq\tfrac{3}{4}$. For these models, SFD does yield a  \emph{`thermodynamical' UV/IR symmetry} which maps high into low temperatures and pressures, and vice-versa.
It is important to note that, independently of their complicated form (\ref{TT}), the $\til T(T)$ transformations  (for $\delta\neq \tfrac{1}{2}$)   can be identified as highly non-trivial realizations of the group $Z_2$,  which  contain the standard simple rule of inversion $\til T=\tfrac{T^2_*}{T}$ as a special case corresponding to $\delta = 1$.  The $Z_2$ nature of the SFD temperature transformations becomes in fact evident if one instead consider their equivalent ``Weyl form",  given by  Eq. (\ref{temptrans}).  

The above established restrictions on the validity of the thermodynamically extended  UV/IR version of SFD symmetry, when combined with  the 
%thermo- and hydro-dynamical
 consistency conditions derived in Sect.\ref{SectSFDfluidThermo}, confirm the  physical (and cosmological) relevance of  a particular sub-family $\delta\in[\frac{3}{4}, 1]$ of self-dual models (\ref{chaply}).

\section{Self-Duality of Apparent Horizons Thermodynamics}

Asymptotically de Sitter universes develop a cosmological horizon with (asymptotically) constant radius. At a sufficiently late time the fluid inside any constant comoving volume $V_0$ occupies a physical volume $V = V_0 \, a^3$ whose radius is greater than the horizon,  which leads to unavoidable causality inconsistencies. 
%in their thermodynamical  description. 
%it becomes unclear what is the meaning of describing  thermodynamically, as whole, such an acausal volume.
On the other hand, the horizons themselves are known to possess proper thermodynamical characteristics \cite{bekenstein,hawking}. 
We shall be considering here the \emph{apparent horizon},\footnote{In general, an apparent horizon may be defined as a marginally anti-trapped surface \cite{hayward1994general}. In the flat FRW universe this coincides with the Hubble radius where space-time expansion becomes superluminal and the expansion of inward radial null geodesics vanishes.} whose physical radius $\Ups_A$, in flat FRW universes, is given simply by the Hubble radius: $\Ups_A = 1 / | H |$. 
Their local definition makes them more appropriate for a thermodynamic description of the \emph{observable universe}
%\footnote{i.e. the matter fluid enclosed within the apparent horizon volume $V_A$.}} 
then the event horizons (see, e.g., \cite{tian2015apparent} for a recent discussion). 

The present section is, accordingly, devoted to the investigation of scale factor self-duality  properties of  the Cai-Kim's  apparent horizon thermodynamics \cite{caikim}. 
The  entropy and temperature prescribed to the apparent horizons  are given by \cite{caikim,cai-akbar}
\br
 T_A \!=\! \frac{1}{2\pi \Ups_A}, \quad S_A \!=\! \frac{1}{4G} (4 \pi \Ups_A^2) \!=\! \frac{8 \pi^2}{H^2}, \quad G \!\equiv\! \frac{1}{8\pi}.\label{aphortd}
 \er
While horizon entropy, $S_A$, follows the standard Bekenstein-Hawking  ``one fourth of area'' rule \cite{hawking}, the  particular choice of the  temperature $T_A$  is partially motivated  by the fact that, with Eq.(\ref{aphortd}),  the Clausius relation $dQ = T_A dS_A $ is equivalent to the Friedmann equations (\ref{frw}) on the horizon, corresponding to an almost adiabatic flux  of energy (heat)  $dQ = - dE$ from the fluid through the apparent horizon. 
If matter satisfies the null energy condition (NEC),  $p+\rho\geq0$, 
then the  $2^{\rm{nd}}$  TD law  holds as well: 
\br
\frac{dS_A}{dt} =\frac{8\pi^2}{|H|^3}(p+\rho)\geq0.\label{aph2law}
\er

It is clear that, by construction, the SFD transformations of all the quantities characterizing the thermal history of the apparent horizon (APH)  can be derived  from the transformation of its  radius:
%Due to the relations $H =\sqrt{\rho/3}=1/\Ups_A$,  the SFD transformations of the corresponding  apparent horizon TD's quantities can be easily obtained from eqs.(\ref{sfd}):
\br
\frac{ \til \Ups_A} {\til a} = \frac{\Ups_A }{a}, \quad\quad {\text{or simply}} \quad\quad  \til r_A =  r_A ,
%\quad\quad \ha a \ha T_A = \ch a \ch T_A ,\quad\quad \ha S_A / \ha a^2 = \ch S_A / \ch a^2 , 
\label{UU}
 \er
where $r_A \equiv \Ups_A / a$ is the comoving radius. They are, in fact, a direct consequence of the fluid density transformations (\ref{sfda}), since  the Friedmann Eq.(\ref{frw}) imposes the relation  $\Ups_A=\sqrt{\rho/3}$ between them. Then the   APH entropy and  temperature  transformations 
%Due to the relations $H =\sqrt{\rho/3}=1/\Ups_A$,  the SFD transformations of the corresponding  apparent horizon TD's quantities can be easily obtained from eqs.(\ref{sfd}):
\br
  \til a \til T_A = a T_A ,\quad\quad\quad  \frac{\til S_A }{\til a^2} = \frac{S_A }{ a^2 } , \label{TSa}
 \er
are obtained by substituting Eqs.(\ref{UU}) into  (\ref{aphortd}). 
The comoving volume of the apparent horizon is not, of course, a  constant. It is instructive to compare what happens with the entropy and energy of the fluid inside this changing volume, in contrast with the transformation (\ref{tdsfd}) for constant $V_0$. 
The physical volumes, $V_A = \tfrac{4}{3} \pi H^{-3}$ and $V = V_0 a^3$, transform quite differently:
\br
\frac{\til V_A}{\til a^3}=\frac{V_A}{a^3},\quad\quad\quad \til V=\frac{v^2_0}{V}.	\label{volumes}
\er
But the internal energy $E_f=\rho V_A$ and  the entropy $S_f=s V_A$ of the fluid inside the apparent horizon turn out to transform in the same way as in Eq.(\ref{tdsfd})
\br
\til S_f=S_f,\quad\quad\quad \frac{\til E_f}{\til a}=\frac{E_f}{a}.\label{av-phv}
\er
Recall that $s = \cal S / a^3$. Now, because of the change of the comoving volume of the apparent horizon, the time evolution (obtained from Eqs.(\ref{frw}))  of the fluid entropies $S_f = s \, V_A$ and $S=s \, V_0 a^3$ manifests an important difference:
\br
\frac{dS_f}{dt}=\frac{2\pi}{3}\frac{\mathcal{S} a}{(aH)^4} (\rho+3p) ,\quad{\text{versus}}\quad \frac{dS}{dt}=0 . \label{fluid2law}
\er
Thus the $2^{\rm{ nd}}$  TD law is \emph{violated} for $S_f$ during an accelerated expansion, when $p < - \tfrac{1}{3} \rho$. 
The simple expression\footnote{It represents in fact the Friedmann equation, when the APH temperature definition (\ref{aphortd}) is taken into account.} $\rho=12\pi^2T^2_A$, which is valid for every fluid,
 reveals another important feature of the APH thermodynamics, namely that  $T_A(a)$ and $S_A(a)$ are always monotonic functions. Notice the difference with the fluid's temperature (and pressure), whose (in general) non-monotonic behavior, as shown in Sect.\ref{SectSFDfluidThermo}, causes certain TD instabilities.

When the universe is filled with the modified Chaplygin gas (\ref{chaply}), the definitions (\ref{aphortd}) of horizon entropy and temperature naturally lead to the \emph{self-duality}  of APH thermodynamics   under  the  thermodynamically  extended SFD transformations: 
\br
 \til T^{2\delta}_A - T^{2\delta}_{\Lambda}=\frac{ T^{4\delta}_{\Lambda} }{ T^{2\delta}_A - T^{2\delta}_{\Lambda}}, \quad\quad \til S^{\delta}_A= S^{\delta}_{\Lambda} - S^{\delta}_A , \label{horts}
 \er
 where $T_{\Lambda} \equiv \tfrac{1}{2\pi}\sqrt{\tfrac{\rho_{\Lambda}}{3}}$  and $S_{\Lambda} \equiv \tfrac{24\pi^2}{\rho_{\Lambda}}$.
 As expected,  the temperature is always a decreasing function of $a\in(0,\infty)$ 
\br
\infty\!>\!T_A\geq T_{\Lambda}\ \text{for}\ \delta\!>\!0, \ \text{or}\ T_{\Lambda}\!\geq\! T_A\!\geq\! 0 \ \text{for}\ \delta\!<\!0,\label{Trange}
\er
and the APH entropy is an increasing one:
\br
0\!\leq\! S_A\!\leq\! S_{\Lambda} \ \text{for}\ \delta\!>\!0, \ \text{or}\ S_{\Lambda}\!\leq\! S_A\!<\!\infty\ \text{for}\ \delta\!<\!0 . \label{Srange}
\er
Observe that in the case of singular asymptotically  de Sitter universes (i.e. for $\delta>0$) the \emph{final}  APH temperature $T_{\Lambda}$ and entropy $S_{\Lambda}$  are equal to the well known values of the  de Sitter event horizon ones. In fact, as one can see from their definitions (\ref{aphortd}), the apparent horizon radius in the $a\rightarrow\infty$ limit do coincide with the cosmological event horizon one. Instead,  for $\delta<0$, i.e.  in the case of  eternal universes, the same de Sitter values  $T_{\Lambda}$ and $S_{\Lambda}$   represent now the \emph{initial} APH temperature and entropy, obtained as  $a\rightarrow 0$. In both cases, the  SFD transformation laws (\ref{horts}) exchange the initial and final states temperatures and entropies, thus acting as a  proper $Z_2$ thermodynamical UV/IR symmetry, whose  fixed points
\br
T_{0A}=2^{\tfrac{1}{\delta}} T_{\Lambda},\quad\quad\quad\quad S_{0A}=2^{-\tfrac{1}{\delta}} S_{\Lambda},\label{fixedAPH}
\er 
represent the  values  of APH temperature and entropy at the deceleration-to-acceleration transition, occurring  at the instant of time $\eta_c$. 

%It should be noted that  the above established scale factor duality properties, specific for the apparent horizons thermodynamics of selected  \emph{self-dual post-big-bang} models (\ref{chaply}),  are quite similar (but \emph{not identical}) to the ones of  the pre-big-bang and cyclic SFD symmetric cosmologies, presented in  our recent paper \cite{ds2016scale}.

\section{Concluding remarks}

We shall complete our investigation with a few comments about  the SFD features of the thermodynamics of the  \emph{``observable universe''}, i.e. thermodynamics of both the apparent horizon \emph{and} the fluid contained inside its volume $V_A$, combined. 
 One of the main challenges here is  the validity of the generalized $2^{\rm{ nd}}$ law, 
 i.e. the growth of the \emph{total} entropy  $S_t = S_f+S_A$, where $S_A$ is the area entropy of the horizon and $S_f = \cal{S} V_A$ the entropy of the fluid inside it. 
 %$\tfrac{dS_t}{dt}\geq 0$ for the total entropy  $S_t = S_f+S_A$. 
  As  can be seen from Eqs.(\ref{fluid2law}) and (\ref{aph2law}), due to the decreasing of $S_f$ in the accelerated expansion phase, it is rather difficult to construct  a $\Lambda$CDM-like cosmology satisfying the  generalized $2^{\rm{ nd}}$ law during the entire evolution of the universe.  It turns out, however, that the considered SFD symmetric models (\ref{chaply}) do offer an example. It is the $\delta=3/4$ self-dual model,  provided  that, as $a\rightarrow\infty$,  the asymptotic value $S_t \to S_{\Lambda}$ obeys the bound  
\br
S_{\La}\geq\frac{2\sigma_{sb}}{3\pi^2},\quad{\text{or equivalently}}\quad \rho_{\Lambda}^{1/4}\leq\frac{2\pi \sqrt{3}}{\left(4\sigma_{sb}\right)^{1/4}}	\; ,		\label{Sbound}
\er
 as we have shown in ref.\cite{ds2016scale}. We should also recall that, apart of its simple TD properties described in Sect.\ref{SectSFDfluidThermo}, this very special \emph{SFD symmetric} cosmological   model   almost  repeats the $\La$CDM energy density evolution --- it behaves as radiation at early times and its late time asymptotic is given by  a cosmological constant accompanied by a cold dark matter. 
 % when $\rho \approx \rho_{\Lambda} +\rho_{dm}/a^3$. 

We have shown  that conformal time scale factor self-duality, when implemented as a symmetry principle, imposes a set of thermodynamical restrictions on the EoS parameters of the self-dual fluid and therefore may be used to select cosmological models.%
\footnote{Including pre-big-bang models, where the slightly relaxed condition of ``partial self-duality'' is also useful \cite{ds2016scale}.}
Our final comment concerns the eventual applications of SFD \emph{not} as a symmetry principle, but rather as an efective transformation relating large and small scales of a pair of \emph{different} dual solutions to the Friedmann equations, each one with a \emph{different} fluid. E.g., for $(a, \rho, p)$ being given by flat $\Lambda$CDM with radiation, Eq.(\ref{sfda}) gives as a dual the universe $(\til a, \til \rho, \til p)$ such that
\br
 \rho = \frac{ \rho_r}{ a^4} + \frac{ \rho_d}{ a^3} + \rho_{\Lambda}, \quad\quad   \tilde \rho =  \frac{\tilde \rho_r}{\tilde a^4} + \frac{\tilde \rho_{dw}}{\tilde a} + \tilde\rho_{\Lambda} , \label{LaCDMdaul}
\er
where $\til \rho_{dw} /\til a$ represents the density of a domain walls gas, dual to the dust term $\rho_d / a^3$.
The methods developed here for self-dual cosmologies can be easily extended to the thermodynamics of these pairs of SFD dual (but not \emph{self}-dual) fluids. Thus one might employ the corresponding  UV/IR \emph{duality} transformations  in the description of the high temperature behavior of a given cosmological model in terms of the low temperature data of its dual.

 \vspace{0.3cm}
 
\textbf{Acknowledgments.} ALAL  thanks CAPES (Brazil) for financial support. The research of UCdS is partially supported by FAPES (Brazil).

\bibliographystyle{JHEP}

\bibliography{ReferencesTDsfd}

%\begin{thebibliography}{}
% etc
%\end{thebibliography}

\end{document}